\documentclass[aps, prl, reprint,notitlepage,showpacs,showkeys,superscriptaddress]{revtex4-1}
\usepackage[utf8x]{inputenc}
\usepackage{graphicx}
\usepackage{amsmath}
\usepackage{amsfonts}
\usepackage{amssymb}
\usepackage{braket}
\usepackage{multirow}

\begin{document}

\title{Thermal and electronic fluctuations of flexible adsorbed molecules: Azobenzene on Ag(111)}
\author{R.J.~Maurer}
\email{reinhard.maurer@yale.edu}
\affiliation{Department Chemie, Technische Universit\"at M\"unchen, D-85748 Garching, Germany}
\affiliation{Department of Chemistry, Yale University, New Haven, CT 06520, United States}
\author{W.~Liu}
\affiliation{Fritz-Haber Institut der Max-Planck-Gesellschaft, Faradayweg 4-6, D-14195 Berlin, Germany}
\affiliation{Nano Structural Materials Center, School of Materials Science and Engineering, Nanjing University of Science and Technology, Nanjing 210094, Jiangsu, China}
\author{I.~Poltavsky}
\affiliation{Fritz-Haber Institut der Max-Planck-Gesellschaft, Faradayweg 4-6, D-14195 Berlin, Germany}
\affiliation{Physics and Materials Science Research Unit, University of Luxembourg, L-1511 Luxembourg}
\author{T.~Stecher}
\author{H.~Oberhofer}
\author{K.~Reuter}
\affiliation{Department Chemie, Technische Universit\"at M\"unchen, D-85748 Garching, Germany}
\author{A.~Tkatchenko}
\email{tkatchen@fhi-berlin.mpg.de}
\affiliation{Fritz-Haber Institut der Max-Planck-Gesellschaft, Faradayweg 4-6, D-14195 Berlin, Germany}
\affiliation{Physics and Materials Science Research Unit, University of Luxembourg, L-1511 Luxembourg}

\date{\today}

\begin{abstract}
We investigate the thermal and electronic collective fluctuations that contribute to the finite-temperature adsorption properties of flexible adsorbates on surfaces on the example of the molecular switch azobenzene C$_{12}$H$_{10}$N$_{2}$ on the Ag(111) surface. Using first-principles molecular dynamics simulations we obtain the free energy of adsorption that accurately accounts for entropic contributions, whereas the inclusion of many-body dispersion interactions accounts for the electronic correlations that govern the adsorbate binding. 
We find the adsorbate properties to be strongly entropy-driven, as can be judged by a kinetic molecular desorption prefactor of 10$^{24}$~s$^{-1}$ that largely exceeds previously reported estimates. We relate this effect to sizable fluctuations across structural and electronic observables. Comparison of our calculations to temperature-programmed desorption measurements demonstrates that finite-temperature effects play a dominant role for flexible molecules in contact with polarizable surfaces, and that recently developed first-principles methods offer an optimal tool to reveal novel collective behavior in such complex systems.
\end{abstract}


\maketitle

Complex molecules adsorbed at inorganic surfaces spark interest as basic building blocks in surface nanotechnology and energy materials~\cite{Barth2005}, but also in the context of biocompatibility of biomolecule-metal interfaces and the structure of solid-liquid interfaces~\cite{Schreiber2000}. Considerable effort goes into characterizing the structure, stability, and dynamics of these systems~\cite{Rosei2003}. In fact, with recent advances in experimental characterization techniques~\cite{Tautz2007} and \emph{ab-initio} methods based on density-functional theory (DFT)~\cite{Nilsson2011}, several systems have been well characterized at idealized conditions, \textit{i.e.} low temperature and ultrahigh vacuum. These include planar aromatic molecules, such as benzene~\cite{Liu2015} and 3,4,9,10-perylene-tetracarboxylic acid (PTCDA)~\cite{Hauschild2010, Mercurio2013b, Ruiz2012, Maurer2015} adsorbed on the Ag(111) surface. Both examples represent comparably rigid molecules forming well-ordered overlayer structures.

\begin{figure}
 \centering\includegraphics{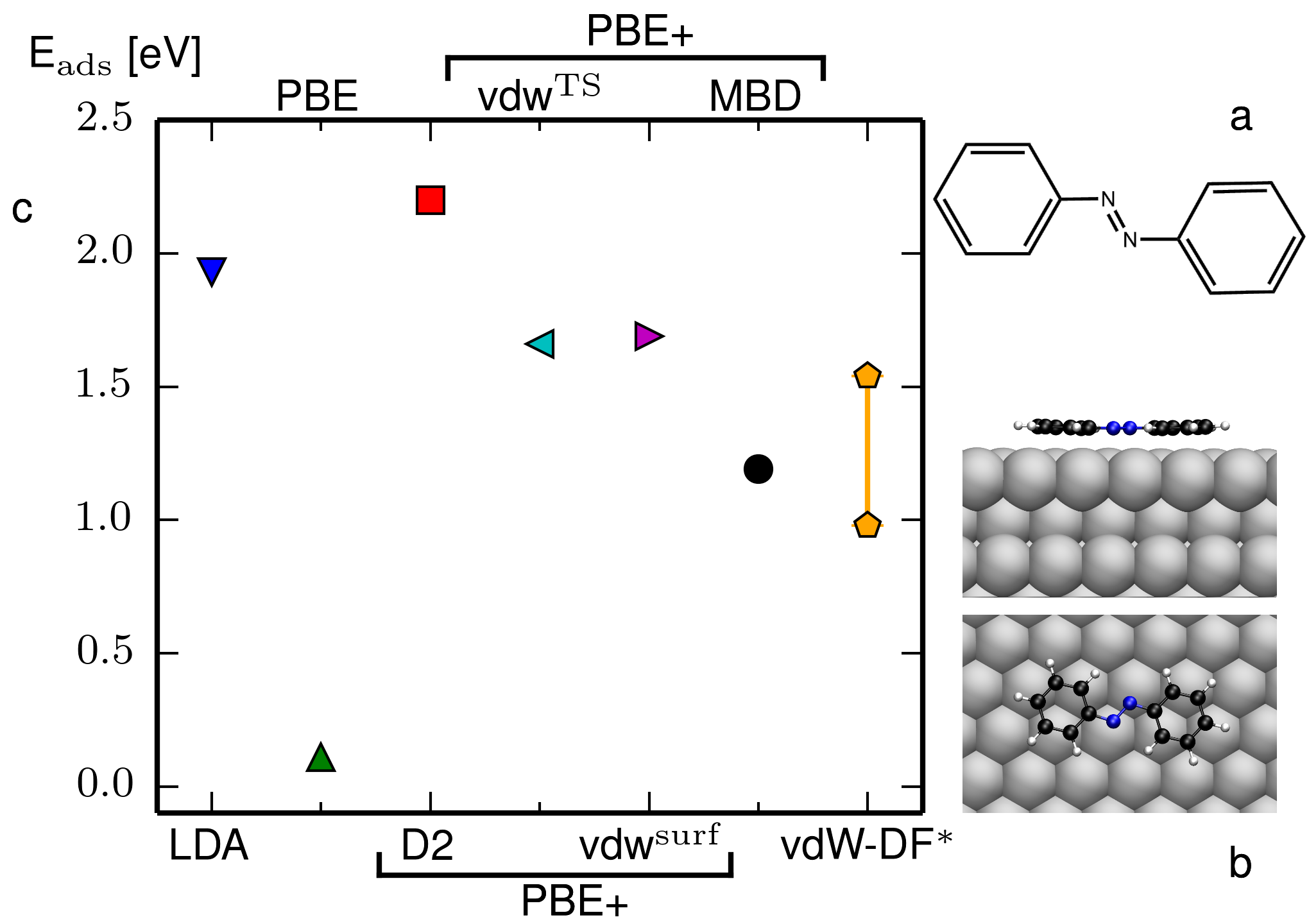}
\caption{\label{fig1} a, Chemical formula of planar azobenzene. b, Side and top view of azobenzene adsorbed at a Ag(111) surface. c, Absolute adsorption energy $E_{\mathrm{ads}}=|\Delta U_{\mathrm{0K}}|$ of azobenzene from Ag(111) as given by different DFT approximations. vdW-DF$^*$ refers to results from vdW-DF methods by Li \emph{et al.}~\cite{Li2012} using different underlying xc-functionals.}
\end{figure}

In contrast, more complex adsorbed systems such as large polymer chains or biological molecules will be neither well-ordered nor rigid. Their flexibility arises from internal torsions and rovibrational coupling in combination with long-range correlations and entails dynamics and reactivity that might be largely shaped by non-trivial thermal and electronic fluctuations.
Whereas the role of thermal fluctuations and corresponding entropic contributions has
always been at the forefront in the modeling of soft condensed matter, their relevance in gas-surface dynamics of flexible molecules in contact with inorganic surfaces is less clear. Long-range correlations induced by electronic fluctuations are an additional complication in the combined molecule/surface system. Several recent works have emphasized the role of temperature for dynamics of benzene on stepped surfaces~\cite{Camarillo-Cisneros2015}, 
conformational switching of porphyrine derivatives on Cu(111)~\cite{Ditze2014, Marbach2014} 
and also for thermal desorption of large alkane chains from graphite~\cite{Paserba2001} and metal 
surfaces~\cite{Weaver2013, Campbell2013a, Fichthorn2002, Fichthorn2007}. Nevertheless, the prevalent view in surface science is that 0~K calculations in the harmonic approximation are often sufficient to reproduce the structure and stability of adsorbed molecules.

To elucidate the possible role of fluctuations, both thermal and induced by electronic correlations, we have chosen to study azobenzene in its planar form (AB) adsorbed on the Ag(111) surface -- a widely used model for on-surface molecular switches (Fig. \ref{fig1}a and b)~\cite{Maurer2012,Mercurio2013}. Azobenzene is a challenging, but also ideal benchmark system containing all relevant features of realistic adsorbates: potential flexibility, low-frequency vibrational modes, and both covalent and dispersion-dominated binding motifs~\cite{Mercurio2010}. Several averaged observables for azobenzene on Ag(111) have been studied experimentally~\cite{Mercurio2010, Schulze2014, Mercurio2014} and theoretically~\cite{McNellis2009, Maurer2012,Mercurio2013,Klimes2012,Berland2015}. The 0~K adsorption energies as predicted by several different DFT approximations range from 0.11 to 2.20~eV (Fig. \ref{fig1}c). A combination of GGA-PBE~\cite{Perdew1996} and the recent many-body dispersion~\cite{Tkatchenko2012, Ambrosetti2014a} (MBD) method yield the closest agreement with the most recent experimental reference of 1.02$\pm$0.06~eV, determined from temperature-programmed desorption (TPD) measurements using the arguably most reliable complete analysis approach~\cite{King1975} based on the Polanyi-Wigner equation~\cite{Schulze2014}. Assuming validity of the experimental analysis, a remaining theoretical/experimental discrepancy to PBE+MBD in adsorption energy of 20\% comes as a surprise, considering the recent quantitative success of this level of theory on the above mentioned examples of benzene and PTCDA on Ag(111)~\cite{Liu2015,Maurer2015}. 

\begin{figure}
 \centering\includegraphics{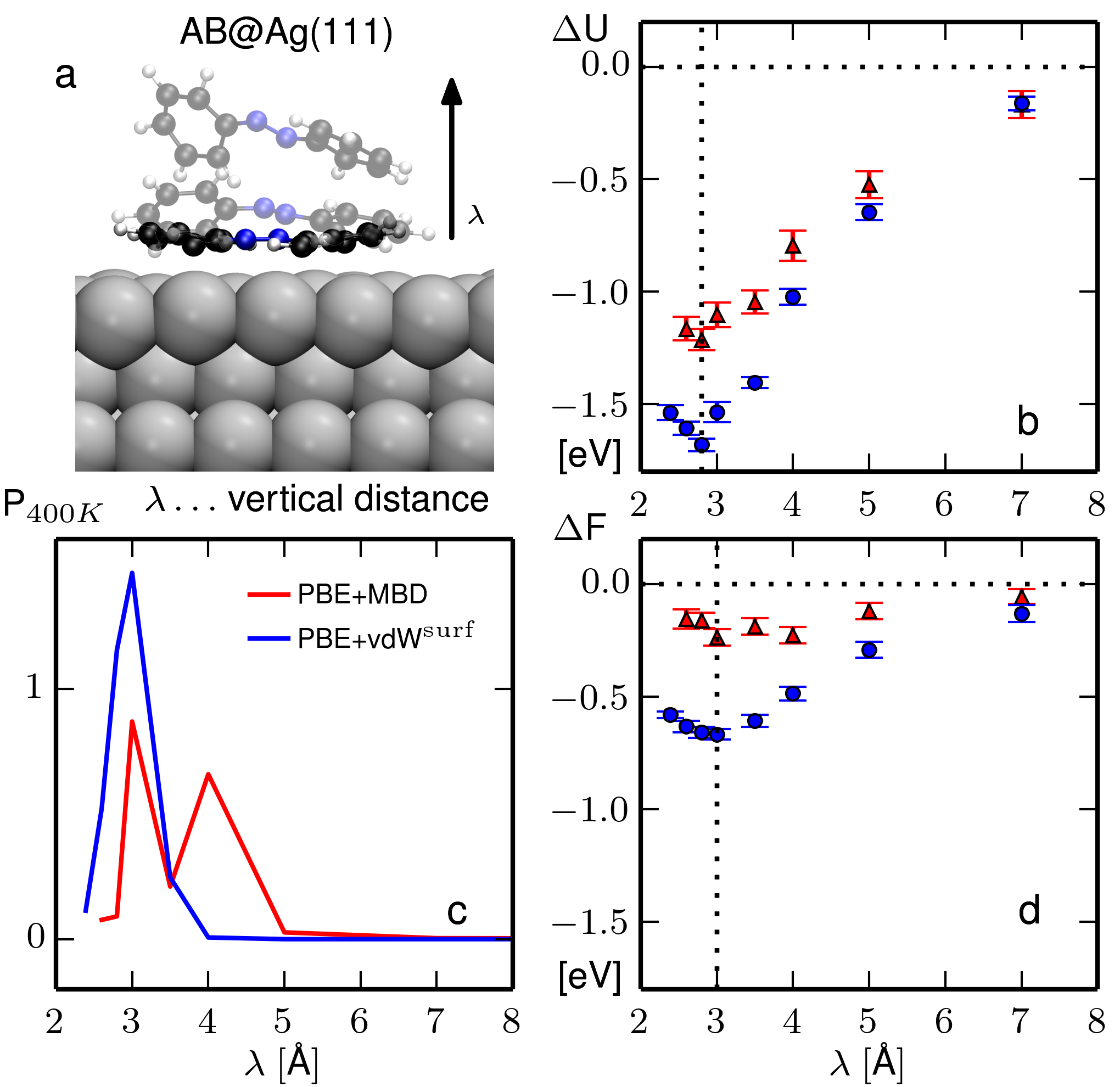}
\caption{\label{fig2} a, A schematic view of the reaction coordinate $\lambda$, which describes the desorption from the surface. b, The average adsorption energy at 400~K as a function of the distance from the surface. c, The probability distribution along $\lambda$ as given by the Boltzmann weight of the free energy. d, The integrated free energy of adsorption (or potential of mean force). Blue squares and lines correspond to PBE+vdW$^{\mathrm{surf}}$, red triangles and lines to PBE+MBD, both are shown with bars indicating the statistical sampling error. }
\end{figure}

To assess the role of thermal fluctuations and their effect on the adsorbate stability, we perform explicit \emph{ab-initio} molecular dynamics (AIMD) simulations of the free energy of adsorption $\Delta F$ for AB on Ag(111) at the experimental desorption temperature of 400~K~\cite{Schulze2014}. 
Hereby thermodynamic integration (\textit{blue moon sampling}) along a chosen reaction coordinate $\lambda$ is performed using constrained AIMD sampling at discrete points along the desorption path:
\begin{equation}\label{eq-termodyn_int}
 \Delta F = \int_{\lambda=2.4\mathrm{\AA}}^{\lambda=14\mathrm{\AA}} d\lambda \left< \frac{\partial H(\lambda)}{\partial\lambda} \right>_{\lambda} .
\end{equation}
The reaction coordinate $\lambda$ in eq. \ref{eq-termodyn_int} has been chosen as the average vertical surface-distance of the two central nitrogen atoms in the AB molecule (\emph{cf.} Fig. \ref{fig2}a) ranging from 2.4 to 14~{\AA}. During AIMD at constant temperature the vertical height of the center of these two atoms has been constrained, whereas all other degrees of freedom were allowed to fluctuate freely in vertical and lateral directions. Simulations have been performed using dispersion-inclusive DFT in form of the PBE+vdW$^{\mathrm{surf}}$ functional~\cite{Perdew1996, Ruiz2012} as implemented in the CASTEP code~\cite{Clark2005} (see SI for more details~\cite{supplemental_refs}). In comparing the differences between the resulting adsorption energy $\Delta U$ (blue circles in Fig. \ref{fig2}b) and the free energy of adsorption $\Delta F$ (blue circles in Fig. \ref{fig2}d) we find a large entropic contribution that reshapes the free energy profile along the reaction coordinate. We calculate finite temperature expectation values by averaging observables along the trajectories and weighting them with the probability distribution as given by the Boltzmann weight of the free energy (blue curve in cf. Fig. \ref{fig2}c). The result is an increased adsorption height of 2.89~{\AA} and a reduced adsorption energy of 1.58~eV when compared to the 0~K results (2.78~{\AA} and 1.68~eV), both in better agreement with x-ray standing wave~\cite{Mercurio2013} and temperature programmed desorption experiments~\cite{Schulze2014} (cf. Table I). 

From the difference in adsorption energy $\Delta U$ and free energy $\Delta F$ we extract an entropy of desorption of 0.24~eV per 100~K, which represents the driving force for this change. Using the Arrhenius equation in the context of transition-state theory we can translate this entropy contribution into a preexponential factor $\nu=(k_BT/h)\cdot\mathrm{exp}(\Delta S/k_B)$ \cite{Tully1984} for desorption of $10^{24}$~s$^{-1}$. This exceeds some of the highest ever reported desorption prefactors ranging up to 10$^{20}$~s$^{-1}$ in the case of desorption of alkane chains~\cite{Paserba2001, Campbell2012, Campbell2013a}. In contrast, a corresponding desorption entropy estimate as given by a simple rigid-rotor harmonic oscillator partition function instead only amounts to 0.12~eV per 100~K, owing to the inability of 0~K calculations to describe the effects of anharmonicity and mode coupling.

\begin{table*}
 \begin{center}
\caption{\label{tab1} Expectation values of adsorption height $\lambda$, the net charge transfer to the molecule $\Delta q$, the dipole moment perpendicular to the surface $\mu_z$, and the adsorption energy $\Delta U$ and entropy $\Delta S$ at the simulation temperature and at 0~K for AB adsorbed at an Ag(111) surface. Brackets denote averages with respect to the probability distribution functions. Errors are given by the standard deviation of the expectation value with respect to the probability distribution.}
\begin{tabular}{ccccccc} 
\hline\hline
 &                &  $\langle\lambda\rangle$ & $\langle\Delta q\rangle$ & $\langle\mu_z\rangle$ &  $|\langle\Delta U\rangle|$ & $|\langle\Delta S\rangle|$ \\
 &   PBE+    & \AA & e & e$\cdot$\AA & eV & eV / 100~K   \\ \hline
\multirow{2}{*}{\rotatebox{90}{0 K}}& vdW$^{\mathrm{surf}}$  & 2.78 & -0.34 & 0.61 & 1.68 & -  \\
				    &    MBD 		     & 2.60 & -0.41 & 0.67 & 1.24 & - \\[2pt] \hline  
\multirow{2}{*}{\rotatebox{90}{400 K}}   & vdW$^{\mathrm{surf}}$  & 2.89$\pm$0.24 & -0.34$\pm$0.05 & 0.59$\pm$0.05 &  1.58$\pm$0.08 & 0.24$\pm$0.02 \\[2pt]
							    & MBD & 3.41$\pm$0.59 & -0.27$\pm$0.06 & 0.48$\pm$0.11 & 0.99$\pm$0.17  & 0.20$\pm$0.04   \\[4pt] \hline
							   & exp. & 2.97$\pm$0.05$^{b}$ & - & - & 1.02$\pm$0.06$^{a}$   \\ \hline\hline 
 \end{tabular} \\
$^a$ TPD measurements of Schulze \emph{et al.}~\cite{Schulze2014} \\ 
$^b$ adsorption height from X-ray standing wave measurements at high coverage and 210~K~\cite{Mercurio2013}
 \end{center}
\end{table*}

Our PBE+vdW$^{\mathrm{surf}}$ AIMD approach correctly captures thermal effects and the sizable entropic contributions originating from averaging over an increasingly larger domain of phase space as the molecule desorbs from the surface. However, at the same time the average adsorption energy of 1.58~eV still exceeds the experimental reference by more than 50\%. This can be traced back to the simple effective account of long-range electronic correlations between molecule and substrate at the PBE+vdW$^{\mathrm{surf}}$ level~\cite{Ruiz2012, Mercurio2013, Liu2013}. Across all distances the combined molecule-surface system is governed by sizable classical and quantum-dynamic fluctuations in the electronic structure. The first are captured in electron density fluctuations throughout the PBE+vdW$^{\mathrm{surf}}$ dynamics. The latter arise from dynamically fluctuating polarizability changes that screen the molecule-surface interaction for a given geometry and are only partly captured at the pairwise-additive vdW$^{\mathrm{surf}}$ level of theory. 

The PBE+MBD (many-body dispersion) approach~\cite{Tkatchenko2012, Ambrosetti2014a, DiStasio2014} as implemented in the FHI-AIMS code~\cite{Blum2009} has recently been shown to accurately capture such long-range correlations and dynamic charge rearrangements for a wide range of extended systems and nanostructures~\cite{Gobre2013,Reilly2013,Ambrosetti2014,DiStasio2014,Liu2015,Maurer2015}. MBD goes well beyond pairwise-additive dispersion schemes by including higher-order many-body contributions and a non-additive geometry-dependence of the polarization response~\cite{Maurer2015}. This approach yields an  adsorption energy of 1.24~eV for AB on Ag(111) at 0~K. Assuming a large overlap between the phase space sampling that is achieved from the PBE+vdW$^{\mathrm{surf}}$ simulations and PBE+MBD we can incorporate dynamic electronic fluctuations into our description via free energy perturbation (\emph{cf.} SI for details)~\cite{Chipot2007}:
\begin{align}\label{eq-FEP}
\Delta F(\mathrm{PBE+MBD}) = \Delta F(\mathrm{PBE+vdW}^{\mathrm{surf}}) - \nonumber \\ 
k_B T [ \ln\braket{\mathrm{exp}(-\beta \Delta E^{\mathrm{vdW}})}_{f} - 
\ln\braket{\mathrm{exp}(-\beta \Delta E^{\mathrm{vdW}} )}_{i}  ] \nonumber \\  
\text{with} \quad   \Delta E^{\mathrm{vdW}}=E(\mathrm{MBD})-  E(\mathrm{vdW}^{\mathrm{surf}}),
\end{align}
where $\beta=1/k_bT$. The resulting MBD-corrected potential energy (red triangles in Fig. \ref{fig2}b) and free energy curves (red triangles in Fig. \ref{fig2}d) are shifted closer to zero with a remaining free energy desorption barrier of only 0.20~eV. This correctly reflects the onset of desorption at this temperature, which is also evident from the 400~K probability distribution function $P(\lambda)=\mathrm{exp}(-\Delta F(\lambda)/k_bT)$  along the reaction coordinate (see red curve in Fig. \ref{fig2}c). The probability of finding AB at a given adsorption height is significant over a wide range from 2.6 up to 5~{\AA} distance from the surface. This can be seen as a result of the temperature-dependent trade-off between adsorption energy and entropy. The increased probability at larger distances originates from a sudden increase in molecular freedom leading to a wide and shallow basin in the PBE+MBD free energy. The latter will become clearer upon discussion of the geometrical details of the desorption process.
 
With the incorporation of long-range electronic correlations at the PBE+MBD level our AIMD simulations are effectively \emph{en par} with experimental observations. The average PBE+MBD adsorption height at 210~K, which we can estimate from $\Delta U(\lambda)$ and $\Delta S(\lambda)$, 2.85$\pm$0.15~{\AA} is in good agreement with our previous calculation of 2.98~{\AA} and results from x-ray standing wave measurements at 210~K~\cite{Mercurio2013}.
The increased average finite-temperature adsorption height, in turn, yields an adsorption energy at 400~K of 0.99$\pm$0.17~eV (see Table I) that is in remarkable agreement with the adsorption energy as extracted from TPD measurements~\cite{Schulze2014}.

We can conclude at this stage that a reliable description of the desorption process requires the interplay of an accurate \emph{ab-initio} electronic structure description and the explicit inclusion of the real-time dynamics of the process yielding the right change in energy and entropy.
To further reiterate the importance of the interplay between energy and entropy we can define an estimate for the desorption temperature as  the ratio between adsorption energy and entropy: $T_{\mathrm{des}}=\Delta U/\Delta S$. As the corresponding PBE+MBD desorption temperature we find 495$\pm$99~K. A description of the energetics or finite temperature effects at a lower level destroys the fair agreement with experiment. For example neglecting many-body dispersion contributions at the level of pairwise dispersion (vdW$^{\mathrm{surf}}$) overestimates the adsorption energy by more than 50\% resulting in desorption temperatures beyond 600~K. Neglecting the real-time dynamics and applying the harmonic approximation results in an  underestimation of the adsorption entropy and desorption temperatures beyond 1000~K.

\begin{figure}
 \centering\includegraphics{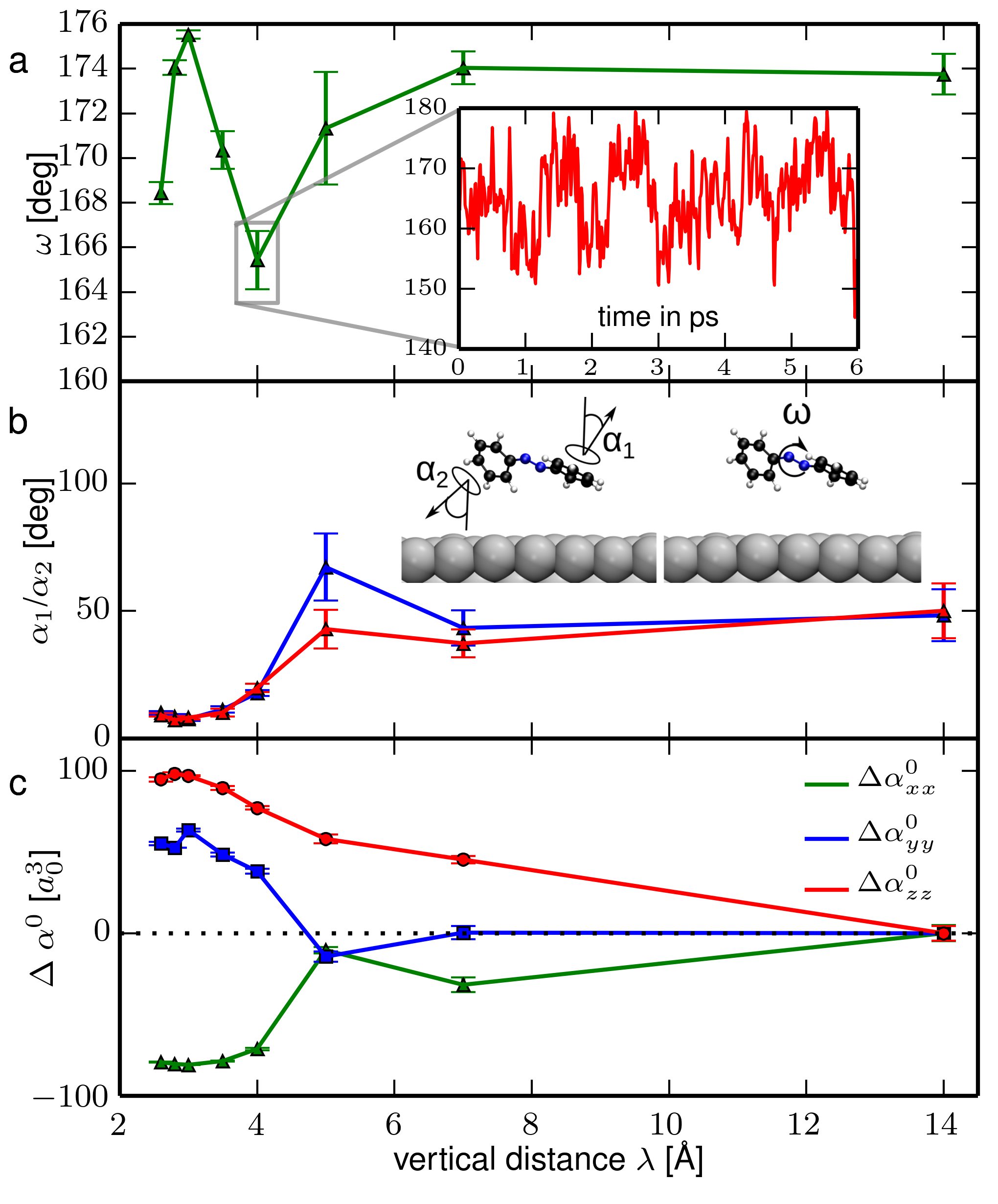}
\caption{\label{fig3} a, Average central CNNC dihedral angle $\omega$ as a function of the reaction coordinate $\lambda$. Also shown as inset is the real-time dynamics of $\omega$ at 4~{\AA} surface distance. b, Average angles $\alpha_1$ and $\alpha_2$ between the normal formed by the two phenyl rings and the surface normal. c, Change in Cartesian static polarizability components. All data points are given with statistical variances shown as error bars.}
\end{figure}

The intricate interplay of both thermal and electronic fluctuations becomes evident from a more detailed analysis of observables such as the molecular geometry along the reaction coordinate~\cite{supplemental_video}. Fig. \ref{fig3}a shows the central dihedral angle $\omega$ and the orientation of the phenyl rings with respect to the surface (angles $\alpha_1$ and $\alpha_2$)  as a function of distance $\lambda$. Close to the equilibrium distance of 3~{\AA} the molecule is almost planar ($\omega\approx$180$^{\circ}$ and $\alpha_{1(2)}\approx$0$^{\circ}$). At intermediate distances between 3 and 7~{\AA} molecular rotations and translations become increasingly accessible and geometric parameters fluctuate wildly as the molecule samples the available conformations. As a consequence of an efficient coupling between vibrational and rotational degrees of freedom, starting at a distance of 4~{\AA}, molecular flapping motion of AB (see inset of Fig. \ref{fig3}a) induces twisting, bending, and subsequently molecular rotations, which turn into free molecular motion beyond 7~{\AA} distance from the surface. The sudden increase in configurational freedom becomes apparent in the broad probability distribution (see Fig. \ref{fig2}c). As a result, all measurable expectation values will be averages over a wide range of surface distances and configurations and subject to equally large fluctuations -- an observation that may be accessible in future experiments from time-resolved single molecule studies or detailed experimental error analysis.

These geometrical fluctuations directly translate to fluctuations in observables derived from the electron density such as the charge transfer between molecule and surface $\Delta q$ and the molecular dipole moment perpendicular to the surface $\mu_z$ (see Table I and Fig. S1 in the Supplemental Material). Strong fluctuations in the adsorption height of the molecule  shift molecular resonances with respect to the Fermi level, which in turn control the charge-transfer between molecule and surface. 
The relevance of an accurate treatment of polarization effects is visible in the Cartesian components of the static molecular polarizability as a function of molecule-surface distance (Fig. \ref{fig3}c). Polarizability components change anisotropically over several orders of magnitude across the reaction coordinate. The individual variances shown as error bars in Fig. \ref{fig3}c do not seem large, however relative to the absolute polarizability of the gas-phase molecule they amount to fluctuations of up to 7\%. The above polarizability changes may serve as a sensitive probe of electronic fluctuations that appear as homogeneous temperature-dependent broadenings in single-molecule surface-enhanced Raman experiments~\cite{Artur2011,Stiles2008,Pettinger2012}.

In summary, we have presented full \emph{ab-initio} molecular dynamics simulations of the desorption of azobenzene from a  Ag(111) surface close to room temperature. For this system, a correct description of adsorption energy and entropy could only be achieved by explicitly accounting for finite-temperature fluctuations and long-range electronic correlations. Whereas the first activate and couple anharmonic modes, the latter screen the dispersion interactions between adsorbate and substrate at all distances. Only recently has it been stated that adsorption entropies have been systematically underestimated in the past~\cite{Campbell2012, Weaver2013} and that processes on surfaces can be purely driven by entropy~\cite{Ditze2014, Marbach2014}. The here calculated desorption entropy of azobenzene on Ag(111) appears to be among the highest ever reported~\cite{Paserba2001, Campbell2013a}. The resulting strong temperature dependence of the vertical adsorption height and other molecular and electronic observables for a comparably small molecule such as azobenzene suggests that finite-temperature effects beyond the harmonic regime are relevant for all but the most rigid adsorbates. In order to approach more realistic model systems in the future and to move toward ambient conditions more research is necessary focusing on the nature and extent of finite-temperature effects, potentially also targeting the time-resolved analysis of single molecule adsorption events. 

Support from the DFG, the European Research Council (ERC-StG VDW-CMAT) and the DoE - Basic Energy Sciences grant no. DE-FG02-05ER15677 is acknowledged for this work. The authors furthermore acknowledge computing time granted by the Leibniz Rechenzentrum under grant no. pr63ya. 


%


\end{document}